\documentclass{aastex61}

\begin{document}
\title{Triggering mechanism and material transfer of a failed solar filament eruption}
\email{yanxl@ynao.ac.cn}
\author{Xiaoli Yan}
\affiliation{Yunnan Observatories, Chinese Academy of Sciences, Kunming 650216, People's Republic of China.}
\affiliation{Sate Key Laboratory of Space Weather, Chinese Academy of Sciences, Beijing 100190, People's Republic of China.}
\affiliation{Center for Astronomical Mega-Science, Chinese Academy of Sciences, 20A Datun Road, Chaoyang District, Beijing, 100012, People's Republic of China.}

\author{Zhike Xue}
\affiliation{Yunnan Observatories, Chinese Academy of Sciences, Kunming 650216, People's Republic of China.}
\affiliation{Center for Astronomical Mega-Science, Chinese Academy of Sciences, 20A Datun Road, Chaoyang District, Beijing, 100012, People's Republic of China.}

\author{Xin Cheng}
\affiliation{School of Astronomy and Space Science, Nanjing University, Nanjing 210093, People's Republic of China}

\author{Jun Zhang}
\affiliation{CAS Key Laboratory of Solar Activity, National Astronomical Observatories, Chinese Academy of Sciences, Beijing 100101, People's Republic of China}

\author{Jincheng Wang}
\affiliation{Yunnan Observatories, Chinese Academy of Sciences, Kunming 650216, People's Republic of China.}
\affiliation{Center for Astronomical Mega-Science, Chinese Academy of Sciences, 20A Datun Road, Chaoyang District, Beijing, 100012, People's Republic of China.}

\author{Defang Kong}
\affiliation{Yunnan Observatories, Chinese Academy of Sciences, Kunming 650216, People's Republic of China.}
\affiliation{Center for Astronomical Mega-Science, Chinese Academy of Sciences, 20A Datun Road, Chaoyang District, Beijing, 100012, People's Republic of China.}

\author{Liheng Yang}
\affiliation{Yunnan Observatories, Chinese Academy of Sciences, Kunming 650216, People's Republic of China.}
\affiliation{Center for Astronomical Mega-Science, Chinese Academy of Sciences, 20A Datun Road, Chaoyang District, Beijing, 100012, People's Republic of China.}

\author{Guorong Chen}
\affiliation{Yunnan Observatories, Chinese Academy of Sciences, Kunming 650216, People's Republic of China.}
\affiliation{Center for Astronomical Mega-Science, Chinese Academy of Sciences, 20A Datun Road, Chaoyang District, Beijing, 100012, People's Republic of China.}

\author{Xueshang Feng}
\affiliation{Sate Key Laboratory of Space Weather, Chinese Academy of Sciences, Beijing 100190, People's Republic of China.}

\begin{abstract}
Soar filament eruptions are often associated with solar flares and coronal mass ejections (CMEs), which are the major impacts on space weather. However, the fine structures and the trigger mechanisms of solar filaments are still unclear. To address these issues, we studied a failed solar active-region filament eruption associated with a C-class flare by using high-resolution H$\alpha$ images from the New Vacuum Solar Telescope (NVST), supplemented by EUV observations of the Solar Dynamical Observatory (SDO). Before the filament eruption, a small bi-pole magnetic field emerged below the filament. And then magnetic reconnection between the filament and the emerging bi-pole magnetic field triggered the filament eruption. During the filament eruption, the untwisting motion of the filament can be clearly traced by the eruptive threads. Moreover, the foot-points of the eruptive threads are determined by tracing the descending filament materials. Note that the filament twisted structure and the right part of the eruptive filament threads cannot be seen before the filament eruption. These eruptive threads at the right part of the filament are found to be rooting in the weak negative polarities near the main negative sunspot. Moreover, a new filament formed in the filament channel due to material injection from the eruptive filament. The above observations and the potential field extrapolations are inclined to support that the filament materials were transferred into the overlying magnetic loops and the nearby filament channel by magnetic reconnection. These observations shed light on better understanding on the complexity of filament eruptions.

\end{abstract}


\section{Introduction}\label{sec:introduction}

Solar filaments with relatively cool and dense materials are suspended in the solar chromosphere and corona (Engvold 1976; Demoulin 1998). Due to absorbing photospheric radiation, solar filaments appear as dark structures on the solar disk in H$\alpha$ center line observation. When solar filaments are born in active-regions, they are called active-region filaments. Usually, active-region filaments often form along the polarity inversion line (PIL) in filament channel (Martin et al. 1998; Gaizauskas 1998). 

For magnetic structure of solar filaments, there are still two points of view (Mackay et al. 2010): One is sheared arcade (Kippenhahn, \& Schl{\"u}ter 1957; Antiochos et al. 1994; DeVore \& Antiochos 2000); The other is flux rope (Kuperus, \& Raadu 1974; Vrsnak et al. 1988; van Ballegooijen \& Martens 1989; Wang, Shi \& Martin 1996; Aulanier et al. 1998; Amari et al. 2000; Gibson et al. 2006; Su \& van Ballegooijen 2013; Yan et al. 2015; Wang et al. 2015; Fan, Gibson, \& Tomczyk 2018; Jiang et al. 2018; Yang et al. 2018). In recent high resolution observation, the untwisting motion was often observed during the eruptions of some active-region filaments (Yan et al. 2014a, b; Yang et al. 2014; Li et al. 2017; Chen et al. 2018). In these cases, the total twist from about 1 $\pi$ to 5 $\pi$ was estimated to be stored in the filament before the eruption. These observations are inclined to support that active-region filaments may have a twisted magnetic structure.

Solar filament eruptions are often associated with solar flares and coronal mass ejections (CMEs) (Gilbert et al. 2000; Gopalswamy et al. 2003; Yan et al. 2011, 2017; Schmieder et al. 2015; Lin 2015; Li et al. 2015; Parenti 2014; Tian et al. 2015; Li et al. 2016; Chen et al. 2018; Zou et al. 2019; Yang \& Chen 2019). Filament eruptions, eruptive flares, and CMEs are believed to be different manifestations in one eruptive process, which usually involves in the disruption of the large-scale magnetic field and restructuring of coronal magnetic field (Forbes 2000; Priest \& Forbes 2002; Lin et al. 2003). Besides, another special type of filament eruption without CME is called failed eruption (Ji et al. 2003; Janvier et al. 2015). In this type of filament eruption, the deformation and the obviously writhe motion of the filament can be often observed (Ji et al. 2003; T{\"o}r{\"o}k, \& Kliem 2010). The associated flare was caused by magnetic reconnection between two legs of the filaments (Alexander, Liu, \& Gilbert 2006; Kliem et al. 2010). 

The triggering mechanisms of solar filament eruptions are variety. Feynman \& Martin (1995) have carried out a statistical work on quiescent filament eruptions and found that the quiescent filament eruptions associated CMEs usually occurred after plenty of new magnetic flux emerged in the vicinity of the filaments. In the following, Chen \& Shibata (2000) have performed two-dimensional magnetohydrodynamic (MHD) numerical simulations and found that appearance of emerging magnetic flux within the filament channel and the outer edge of the filament channel can lead to the rise of the flux rope and a current sheet formation. The above observations and simulations were confirmed by many following works (Jing et al. 2004; Yan et al. 2011; Li et al. 2015; Jiang et al. 2016). Moreover, filament eruptions associated with solar flares and CMEs are found to be closely associated with flux cancellation (Livi et al. 1989; Linker et al. 2003; Sterling et al. 2018). Kink instability and torus instability are suggested to be the trigger of the flux rope eruptions and filament eruptions (T{\"o}r{\"o}k et al. 2004; Kliem \& T{\"o}r{\"o}k 2006; Yan et al. 2014a, b; Guo et al. 2010; Bi et al. 2015; Liu et al. 2016). Besides, the emerging magnetic flux with opposite current to the filament can also trigger the filament eruption (Wang et al. 2016). Interestingly, in some cases, if the filaments have a strong rotation motion during their eruption, even the decay index satisfied with the threshold of torus instability, the filaments still experienced failed eruptions (Zhou et al. 2019). In addition, magnetic reconnection between two groups of magnetic field lines below and above the filaments also plays an important role in the filament eruptions (Moore et al. 2001; Antiochos et al. 1999; Forbes \& Isenberg 1991; Antiochos, DeVore, \& Klimchuk 1999; Lin \& Forbes 2000; Sterling et al. 2004; Chen et al. 2016; Liu et al. 2018b). Song et al. (2013) found that some fast CMEs associated with the eruptions of the well-developed long quiescent filaments were not accompanied by any flares. They suggested that ideal macroscopic magnetohydrodynamic instability plays an important role in accelerating CMEs in quiet-Sun regions with weak magnetic field. Therefore, how filament eruptions are initiated is still unpredictable for a specific event.

The classical standard flare model, called as CSHKP flare model, has been proposed for several decades (Carmichael 1964; Sturrock 1966; Hirayama 1974; Kopp \& Pneuman 1976). This model proposes that magnetic reconnection between two groups of overlying magnetic field lines with opposite directions occurs under a flux rope/a filament. A current sheet forms between the rising flux rope/the filament and the newly formed hot cusp-shaped coronal arcade during solar flares. At the same time, two bright ribbons form at either side of PIL in the chromosphere. The ribbons located at the feet of the overlying magnetic loops and separate from each other during solar flares. In CSHKP model, the eruption of a prominence or a flux rope should be accompanied by a CME and a flare. The overlying magnetic loops become the flux of the filament/the flux rope via magnetic reconnection and eject into the interplanetary space. Especially, the event occurring on 2017 September 10 at the solar limb studied by Yan et al. (2018), Liu et al. (2018a), Shen et al. (2018); Liu et al. (2019b), and Cheng et al. (2019) is very consistent with standard CSHKP model and catastrophic loss of equilibrium model (Forbes \& Isenberg 1991; Lin \& Forbes 2000). 

Though there are many structures of solar filaments seen in the recent high resolution observations, the explanations on the fine structures and eruption process of solar filaments are still controversial. In this paper, we present the fine structure of an active-region filament and detailed eruption process associated with a C6.7 flare, which were revealed by the NVST and SDO observations. This paper is organized as follows: the observations are described in Section \ref{sec:observations}. The details of results are presented in Section \ref{sec:results}. The conclusion and discussion are given in Section \ref{sec:conclusion}.

\section{Observations}\label{sec:observations}

The NVST is the biggest ground-based solar telescope with a 986 mm clear aperture in China, which is located in the Fuxian Solar Observatory of Yunnan Observatories, Chinese Academy of Sciences (CAS). It is designed to observe the fine structures in the Sun and their activities in the multiple atmospheric layers with high spatial resolution (0.$^\prime$$^\prime$165 per pixel in the chromosphere, 0.$^\prime$$^\prime$04 per pixel in the photosphere) and high temporal resolution ($\sim$12 s) (Liu et al. 2014; Yan et al. 2019). In this paper, the TiO and the H$\alpha$ line-center images are acquired at 7058 ~{\AA} and  6562.8~{\AA}, respectively. Moreover, off-band H$\alpha$ images acquired at $\pm$ 0.5 \AA\ of the line center are used to show the blue-shift and red-shift Doppler velocity during the filament eruption. The data were calibrated from Level 0 to Level 1 with the dark current subtracted and flat-field corrected. To obtain better quality images, the calibrated images were reconstructed by using the speckle masking method from Level 1 to Level 1 + by Weigelt (1977), Lohmann et al. (1983), and Xiang et al. (2016). These data can be accessed at \url{http://fso.ynao.ac.cn/dataarchive_ql.aspx}. 

The EUV 304 \AA, 171 \AA, and 131 \AA\ images obtained by the Atmospheric Imaging Assembly (AIA; Lemen et al. 2012) on board the Solar Dynamic Observatory (SDO; Scherrer et al. 2012) with a spatial resolution of $1.^{\prime\prime}5$ and a cadence of 12 s are used to display the process of the filament eruption. The 304 \AA\ images obtained by Extreme Ultraviolet Imager (EUVI; Howard et al. 2008) telescope on board STEREO-A are also employed to show the filament eruption from another visual angle. They have a spatial resolution of $1.^{\prime\prime}6$ and a cadence of 10 minutes. Photospheric line-of-sight (LOS) magnetograms and the continuum intensity images with a 45 s cadence and a pixel scale of 0.5$^\prime$$^\prime$ are provided by the Helioseismic and Magnetic Imager (HMI;  Schou et al. 2012) on board SDO. 

\section{Results}\label{sec:results}
\subsection{Active region NOAA 12740 and an active-region filament.}
On May 9, 2019, active region (AR) NOAA 12740 appeared near the disk center. Figures 1a and 1b show appearance of the AR in TiO image observed by the NVST and the line-of-sight magnetogram observed by the SDO/HMI. The AR is composed of a main leading negative polarity and  a scatter following positive polarity. There are also some scattered positive polarities near the north of the leading sunspot. Figures 1c and 1d show appearance of an active-region filament situated at the northeast of the leading sunspot in 304 \AA\ image observed by AIA/SDO and H$\alpha$ image observed by the NVST. The filament exhibited an M-shaped configuration (the white arrows pointing to the filament in Figs. 1c and 1d). One end of the filament located the leading negative sunspot and the other end situated in the concentrate small positive polarity (see Fig. 3a). Seen from H$\alpha$ image in Fig.~1d, many thin threads along the filament spine make up its main structure. 

\subsection{Emergence of a small-scale bi-pole magnetic field under the filament before the filament eruption.}
Before the filament eruption, a brightening in 304 \AA\ images first appeared near the left part of the filament marked by the blue dashed box in Fig.~2a. The time evolution of this region in 304 \AA\ images can be seen from the sub-images at the top of Fig.~2a. Following AR evolution, the brightening region became larger and larger. The green line in Fig. 2c shows the evolution of the normalized intensity of this region. The red and the blue lines in Fig. 2c show the GOES soft X-ray profile at 0.5-4 \AA\ and 1-8 \AA\ bands. There was a C6.7 solar flare occurring in this AR. The flare started at 05:40 UT, peaked at 05:51 UT, and ended at 05:56 UT. According to the 304 \AA\ observation and the normalized intensity plot in green line in Fig. 2c, the onset of the brightening appeared at about 04:38 UT and lasted until the ending of solar flare C6.7 (see the animated version of Fig.~2). It is obvious that the normalized intensity increased slowly before the onset of the flare and then reached the maximum value before the peak of the flare. Of course, the intensity was affected by the flare. By checking the evolution of the LOS magnetograms, the emergence of bi-pole magnetic field marked by the blue dashed box in Fig. 2b occurred in the photosphere. The three sub-images of the line-of-sight magnetograms at the top of Fig. 2b show the emerging process of the bi-pole magnetic field. At 01:08:50 UT, there were several small mixed magnetic polarities in the middle of the blue dashed box. Next, the positive and the negative polarities emerged here. The bi-pole magnetic field separated from each other during the process of the emergence. The negative ones moved to the north and the positive one moved to the south. Note that the red and the blue arrows in the sub-images of Fig. 2b indicate the positive and the negative polarities of the emerging bi-pole. The change of the positive and the negative magnetic flux is shown in Fig. 2d. The red and the blue lines denote the positive and the negative magnetic flux in the region marked by the blue dashed box in Fig. 2b. The positive and the negative magnetic flux increased at the same time. From the spatial position and time sequence,  the occurrence of the brightening and the C-class flare were closely related to the emergence of the bi-pole magnetic field. It implies that the emerging magnetic fields reconnected with the magnetic structure of the filament system and resulted in the brightening and the instability of the filament.

\subsection{Transfer of solar filament materials by magnetic reconnection}

Figure 3 shows the eruption process of the active-region filament associated with a C-class flare acquired in NVST H$\alpha$ observation with the line-of-sight magnetograms superimposed. The contour values of magnetic fields are $\pm$ 100 G. The blue and the red contours respect the negative and the positive polarities. It is obvious that the left foot-point of the filament is rooting in a concentrated positive polarity while the right foot-point seems to be rooting in the penumbra of the main sunspot with negative polarity (see Fig.~3a). The filament almost located along the polarity inversion line (PIL). The north region of the filament exhibits positive polarity and the south region of the filament exhibits negative polarity. Therefore, the chirality of the filament belongs to the sinistral filament (Martin 1998).

Before the obvious rising of the filament, flare ribbons first appeared at the left part of the filament (see Fig.~3b). The brightening region became larger following the rising of the filament. Subsequently, the left part of the filament experienced a rapid rising. The threads of the filament can be seen to be rotating clockwise (see Figs. 3c-3g). At the original stage of the filament eruption, the filament also exhibited a twisted structure by tracing the threads along the spine of the filament (see the animation in Fig.~3 from 05:48 UT to 05:52 UT). The observations are inclined to support that the magnetic structure of the filament is a flux rope. The rising of the filament was from the left to the right. Interestingly, following the rising of the filament main body, most the eruptive threads of the filament were found to be connecting the weak magnetic fields near the main negative sunspot (see the green arrows in Figs. 3d-3g). Moreover, the brightening of these foot-points was caused by the falling filament materials. However, these foot-points of the threads can not be seen before the filament eruption. Therefore, the possible explanation is that the filament materials were transferred into the overlying magnetic loops by magnetic reconnection between the filament magnetic structures and the overlying magnetic fields. The rightmost part of the filament connecting with the main negative sunspot did not erupt (Figs. 3h-3i). At the final stage of the filament eruption, a group of the filament threads can be seen to be rooting in a very weak negative magnetic fields pointed by the green arrows in Figs. 3h-3i. 

After the eruption of the filament, the left part of the filament changed its connection and extended to the northward. It seems that a new filament formed. One end of its spine extended to the north and the other end was still connecting the penumbra of the large negative sunspot as the original filament does. The newly formed small chromospheric fibrils are between the original left foot-point of the filament and the negative leading sunspot (see Figs. 3f-3i). When the filament erupted, the brightening in the positive polarity signed by the yellow arrows and in the negative polarity signed by the green arrows in Figs. 3d-3g appeared at the same time. It implies that the magnetic structures of the filament reconnected with the surrounding overlying magnetic field. 

Figures 4 and 5 show eruption process of the active-region filament in off-band H$\alpha$ images at -0.5 \AA\ and +0.5 \AA\ observed by the NVST with the contours of the line-of-sight magnetograms superimposed. The red and the blue contours represent the positive and the negative magnetic fields. The contour levels of magnetic fields are $\pm$100 G. The green arrows denote the foot-points of the filament threads. The right part of the filament can be seen clearly in H$\alpha$ blue wing observations after the onset of the filament eruption, which is caused by the lift of the filament. The threads of the right part filament are clearly seen to be rooting in the penumbra of the large negative sunspot. During the filament eruption, this part of the filament did not change so much (Figs. 4d-4i). Figures 5a-5c show the H$\alpha$ red wing observations at the beginning of the filament eruption. In the following, many eruptive threads of the filament can be seen to be rooting in the weak negative magnetic fields along the spine of the filament (Figs. 5d-5i). The red wing observation corresponds to the down-flow during the filament eruption. The foot-points of the eruptive filament threads in the red wing images are much easier to be identified than those in the blue wing images. The most of the filament materials fell down along the threads that rooted in the weak negative polarity signed by the green arrows in Figs. 5d-5i. 

The process of the filament eruption was covered by the observation of SDO/AIA. Figure 6 shows appearance of the brightening at the right part of the spine of the filament. When the filament rose up from the east to the west, the filament began to interact with the overlying magnetic loops. The overlying magnetic loops can first be seen in EUV images at about 05:45 UT (see the animated version of Figure~6) before the rising of the filament. At 05:55 UT, the brightening began to appear at the position marked by the blue arrows in Figs. 6a-6c. The red arrow in Fig. 6c, 6f and 6i denotes the overlying magnetic loops seen in the 131 \AA\ images. The brightening expended from the right side to the left side of the filament spine accompanying with the filament eruption (see the dotted blue arrows in Figs. 6d-6i). Moreover, the brightening material of the filament can be seen to fall down along the eruptive threads from the brightening region to the foot-points of the eruptive filament threads (see the Figure~6 animation). Note that the brightening was not observed in H$\alpha$ center and off-band images. That is to say, the reconnection between the structure of the filament and the overlying fields may occur at the higher atmosphere. When the filament material transferred into the overlying loops, the obvious down-flow can be observed in the H$\alpha$ observation. This observation evidence implies that the magnetic structure of the filament reconnected with the overlying magnetic loops and the materials of the filament were transferred into the overlying magnetic loops.

During the filament eruption, the obvious brightening appeared at the right part of the filament spine. Figure 7a shows the normalized intensity in the region outlined by the box in Figure 6a. The brightening began at about 05:56 UT. The intensity of three EUV lines all increased and reached the peak at about 06:00 UT. Moreover, the velocity of the down-flows from the site of the brightening along the green dotted line in Figure 6a was measured by using time-distance diagrams. There are three time-distance diagrams shown in Figures 7b-7d to be used to evaluate the velocity of the down-flow by tracing the change of the intensity. The projection velocities of the discontinuous down-flows range from 127 to 155 km/s.

To address why a new filament appeared at the northwest of the filament, the 304 \AA\ images observed by the STEREO were used to show the process of the filament eruption from another visual angle. Before the filament eruption, there is a large quiescent filament at the northwest of the filament marked by the green arrow in Fig. 8a. From the 304 \AA\ image in Fig. 1c, it can be also seen that there is a filament channel. The blue arrow in Fig. 8b marked the erupting filament. When the filament erupted toward the northwest of the filament, the structure of the filament interacted with the filament channel, the materials of the filament can be seen to drop from the filament and inject into the filament channel (see the blue arrows in Fig. 8c and 8d). When the filament channel captured these materials, a new filament appeared in H$\alpha$ image. That is why a new filament appeared at the northwest of the filament. The AIA EUV observations can be seen from Supplementary Movies~3.

By checking the SOHO/LASCO observation, it is found that this filament eruption was not associated with a CME. The failed eruption process of the filament can be seen from the 304 \AA\ images in Figures 8a-8c. Most of the filament material fell down along the filament threads. Therefore, the filament eruption experienced failed eruption. 

To show the rising velocity and untwisting motion of the filament during its eruption, two types of time-distance diagrams were made along the slit of AB and along the slit of CD. The slits are marked in Fig. 9c. Figures 9d-9f show the time-distance diagrams along the silt of AB and CD, respectively. The time-distance diagrams of Figs. 9d and 9e are made by using 171 \AA\ and 304 \AA\ images along the silt of AB. The rising projected velocity of the filament was calculated to be 166 km $s^{-1}$ by tracing the filament structures. During the downward motion of the filament structure, there is an oscillation outlined by the green lines in Figs. 9d and 9e. The slit of CD is perpendicular to the spine of the filament. From the time-distance diagrams of Fig. 9f, the threads of the filament can be seen to be rotating from one side to the other of the filament. The blue dotted lines denote the motion of the filament threads. It demonstrates that the untwisting motion occurred during the filament eruption. From 05:50 UT to 05:59 UT, the filament threads rotated semi-circle. From 06:00 UT to 06:04 UT in Fig. 9f, the filament threads rotated another semi-circle. Therefore, the twist is estimated to be at least one turn.

\subsection{Potential field extrapolation}

Figure 10 shows the overlying magnetic structure of the filament superimposed on the line-of-sight magnetogram and the H$\alpha$ images by using potential field source surface (PFSS) model, respectively. The yellow and the cyan lines  are the selected potential field lines connecting from the positive polarity at the north to the negative polarity at the south. There are two groups of magnetic loops shown in Figure 10. One group of magnetic loops in yellow indicates the overlying loops from the main negative sunspot, while the other indicates the overlying loops connecting the weak satellite bi-pole (see Figure 10a). These overlying magnetic loops are trapping the filament before its eruption (see Figure 10b). During the filament eruption, the filament materials can be seen to be falling down along the eruptive thread to the foot-points of the overlying magnetic loops. The brightening appeared at the foot-points of these overlying magnetic loops (see Figures 10c and 10d). These observations imply that the filament materials were transferred into the overlying magnetic loops.

\section{Conclusion and discussion}\label{sec:conclusion}

In this paper, we studied the detailed process of a failed filament eruption associated with a C-class solar flare in AR NOAA 12740 on May 9, 2019. We found that the filament eruption was caused by the emergence of magnetic flux that reconnected with the magnetic structure of the filament. Moreover, the obvious untwisting motion of the filament was also observed at the initial phase of its eruption. It implies that the magnetic structure of the filament may be a flux rope. During the filament eruption, the foot-points of the eruptive threads are determined by tracing the down-flow along these threads. It is found that the foot-point distribution of the eruptive threads is dispersive, which are rooting in many small satellite polarities along the spine of the filament. By using extrapolation of potential fields, it is found that these satellite polarities are the foot-points of the overlying magnetic loops. After the filament eruption, a new filament appeared at the north of the original filament. These observations are inclined to support that the filament materials were transferred into the overlying magnetic loops and the filament channel at the north. The release of energy comes from magnetic reconnection between the magnetic structure of the filament and the surrounding overlying magnetic loops. 

The eruption process of solar filaments is variety. In some cases, the structure of solar filaments fully erupt (Wang et al. 2003; Jiang et al. 2009; Huang et al. 2019). However, in other cases, solar filaments experience partial eruption (Gilbert et al. 2007; Zhang et al. 2015; Cheng et al. 2018; Zheng et al. 2019). Some filament eruptions produce the CME and others are not associated with the CME (Shen et al. 2011). To better understand these issues, we need to know the fine structures of solar filaments and the surrounding magnetic topology. Usually, the H$\alpha$ image before the filament eruption is used to determine the foot-points of solar active-region filaments. In this event, the foot-points of the filament were found to be rooting in a concentrated positive polarity  and in the super-penumbra of the main negative sunspot. This filament belongs to the partial eruption due to that its right part did not erupt, which was rooting in the super-penumbra of the main negative sunspot.   

In this event, the eruptive part of the filament exhibited an obvious untwisting motion. It implies that the filament had a twisted magnetic structure before the filament eruption (Kurokawa et al. 1987; Schmieder et al. 1985; Filippov et al. 2015; Chen et al. 2019). The twist in the filament structure is evaluated to be at least one turn by tracing the eruptive threads of the filament. At the beginning of the filament eruption, the brightening first appeared  at the left part of the filament and then the large-scale magnetic loops formed slowly. It implies that the magnetic reconnection occurred between the structure of the filament and the surrounding magnetic loops. During the filament eruption, the filament materials were observed to be injected into the overlying magnetic loops. Therefore, it implies that the main released energy of the flare came from the magnetic reconnection between the magnetic structure of the filament and the surrounding overlying magnetic loops.

In previous studies, the filament eruptions are often found to be associated with magnetic cancellation and magnetic emergence (Romano et al . 2003; Jing et al. 2004; Yan et al. 2011; Hou et al. 2019; Liu et al. 2019a). Chen \& Shibata (2000) carried out a MHD simulation and found that the small-scale magnetic emergence near a flux rope can reconnect with the flux rope to cause the eruption of the flux rope. As the spatial and temporal resolution of the observational data is low in the past, the connection between magnetic cancellation or magnetic emergence and filament eruption cannot identified well. In this event, the emergence of a small-scale bi-pole magnetic polarities is found below the filament and caused the magnetic reconnection between the filament and the emerging bi-pole magnetic field. The exact evidences of magnetic reconnection are as follows: First, the emergence of the bi-pole polarities in the photosphere; Second, the brightening above the site of the magnetic emergence observed in the 304 \AA\ images. The brightening area became larger and larger accompanying with the magnetic emergence and then the filament eruption was following. These observations imply that magnetic reconnection occurred between the filament and the small-scale emerging bi-pole polarities. It is worthy to note that the flux emergence began at 01:00 UT, which is earlier about four hours than the onset of the filament eruption. During the emergence of the bi-pole before 04:00 UT, the transient brightening appeared below the filament. We assume that the instability of the filament caused by the magnetic reconnection between the filament and the emergence flux is a slow process. 

The filament materials often fall down along the spine of the filament during the filament eruption and the brightening appear at the foot-points of filaments. In classical solar eruptive models, the overlying magnetic loops of a filament or flux rope become the toroidal field of the filament or flux rope through magnetic reconnection below the filament or flux rope (Lin \& Forbes 2000) and the post-flare loops form after the filament or the flux rope eruption. In this event, the filament materials were transferred into the overlying magnetic loops and another newly formed filament. After the filament eruption, small groups of post-flare loops appeared below the left part of the filament and no post-flare loops appeared below the middle and the right part of the filament (the region covered by the yellow and the cyan potential field lines in Figure 10). In standard flare model, magnetic reconnection occurs below a magnetic structure (filament or flux rope) and a current sheet forms following the eruptive structure. The post-flare loops appear at the bottom of the current sheet. The event studied in this paper is different from the scenario mentioned above. {\bf The potential explanation} is that the transfer and redistribution of the filament materials were caused by magnetic reconnection between the magnetic structure of the filament and the overlying magnetic fields. Schematic sketch is illustrated in Figure 11 to show the change of the magnetic structures of the filament system before and after the filament eruption. The blue loops indicate the overlying magnetic fields and the red curved magnetic field lines indicate the magnetic structure of the filament before the filament eruption in Figure 11a. When the filament erupted, a part of the magnetic structure reconnected with the overlying magnetic fields. Some magnetic field lines of the filament and overlying magnetic loops changed their connections (see Figure 11b). The filament materials were transferred into the overlying magnetic loops and fell down to their foot-points. This event is different from that studied by Li et al. (2016) and Xue et al. (2016), which found the magnetic reconnection between the filament and nearby coronal loops or chromospheric fibrils. 

To address whether the newly detected thread foot-points have a relation to the magnetic reconnection with the overlying fields, there are some discussions as follows: First, the obvious discontinuous fast down-flows occurred after the brightening appeared at the right part of the filament spine, which implies that where magnetic reconnection happened. In previous studies, we have not found the appearance of the brightening in the filament spine and the materials falling down from the brightening site to the lower atmosphere during the untwisting motion of active-region filament eruption (Yan et al. 2014a, 2014b); Second, since the cool material was injected into the overlying magnetic loops, the relatively low temperature of these loops make them to be observed in H$\alpha$ center images and off-band H$\alpha$ images. Besides, the brightening was only observed in EUV observation. It means that the reconnection occurred at the higher atmosphere. These facts are inclined to support that the magnetic reconnection occurred between the filament and the overlying fields. Note that we cannot rule out absolutely whether these eruptive threads belong to the filament system before the filament eruption due to the low temperature and  the low density. For the newly formed filament at the northwest of the filament, its material came from the erupting filament. During the filament eruption, the filament material was found to be injected into the filament channel. Therefore, this is the first time to observe magnetic reconnection between the magnetic structures of the filament and the filament channel in the high resolution chromospheric observation. These results extend our understanding on the complex process of solar filament eruptions.

\begin{figure}
\plotone{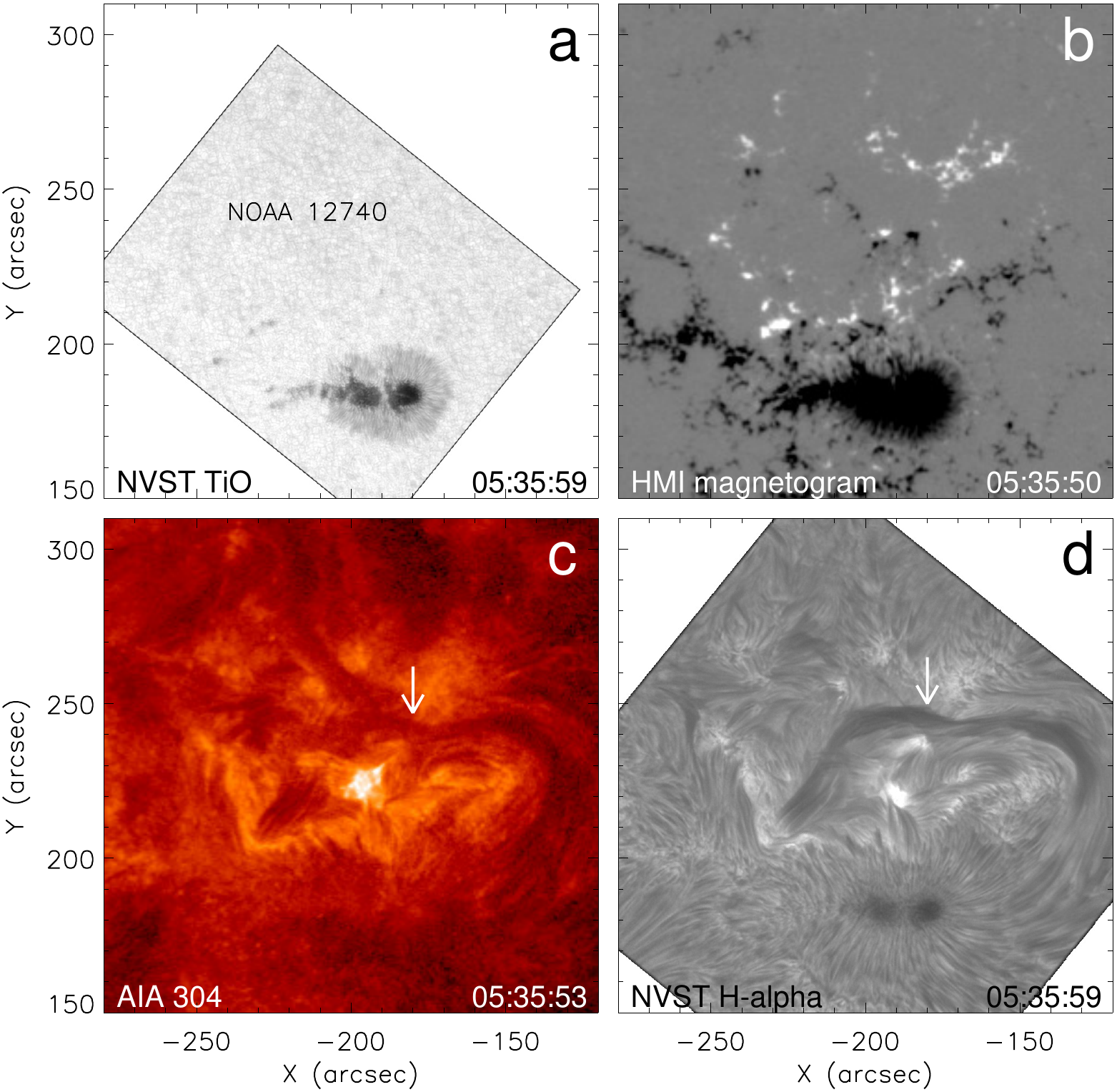}
\caption{Appearance of active region NOAA 12740 and the filament. (a) TiO images observed by the NVST. (b) Line-of-sight magnetogram observed by SDO/HMI. The white and the black patches denote the positive and the negative magnetic polarities, respectively. (c) 304 \AA\ image observed by SDO/AIA. (d) H$\alpha$ images observed by the NVST. The white arrows in Figs. 1c and 1d point to the active-region filament. \label{fig1}}
\end{figure}

\begin{figure*}
  \centering
   \includegraphics[width=12cm]{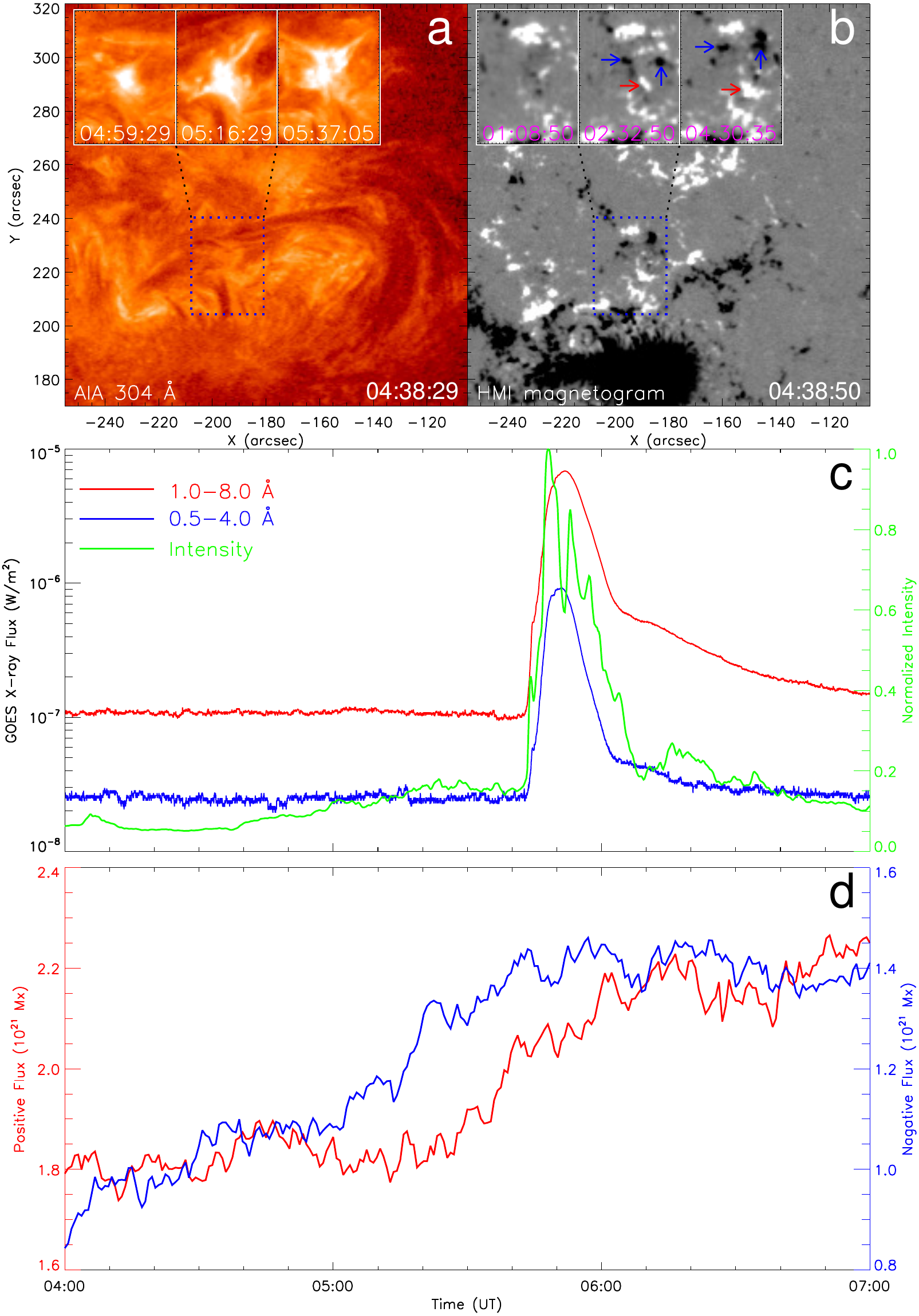}
\caption{Precursors of the active-region filament eruption in 304 \AA\ images and line-of-sight magnetograms. (a) 304 \AA\ image observed by SDO/AIA. The sub-images show the evolution of the brightening at the dashed blue box. \textbf{An animation of this panel is available in the online Journal. The animation includes the wide-field and sub-images, and the images proceed from 3:03 to 7:00~UT}. (b) Line-of-sight magnetogram observed by SDO/HMI. The sub-images show the magnetic field emergence at the dashed blue box.(c) The GOES soft X-ray profile of 0.5 - 4 \AA\ (the blue line) and 1- 8 \AA\ and the normalized intensity (the green line) of the dashed blue box in in Fig. 2a. (d) The evolution of the positive and the negative magnetic flux in the dashed blue box in Fig. 2b. The arrows in Figs. 1c and 1d point to the active-region filament.}
     \label{fig1}
   \end{figure*}

\begin{figure}
\plotone{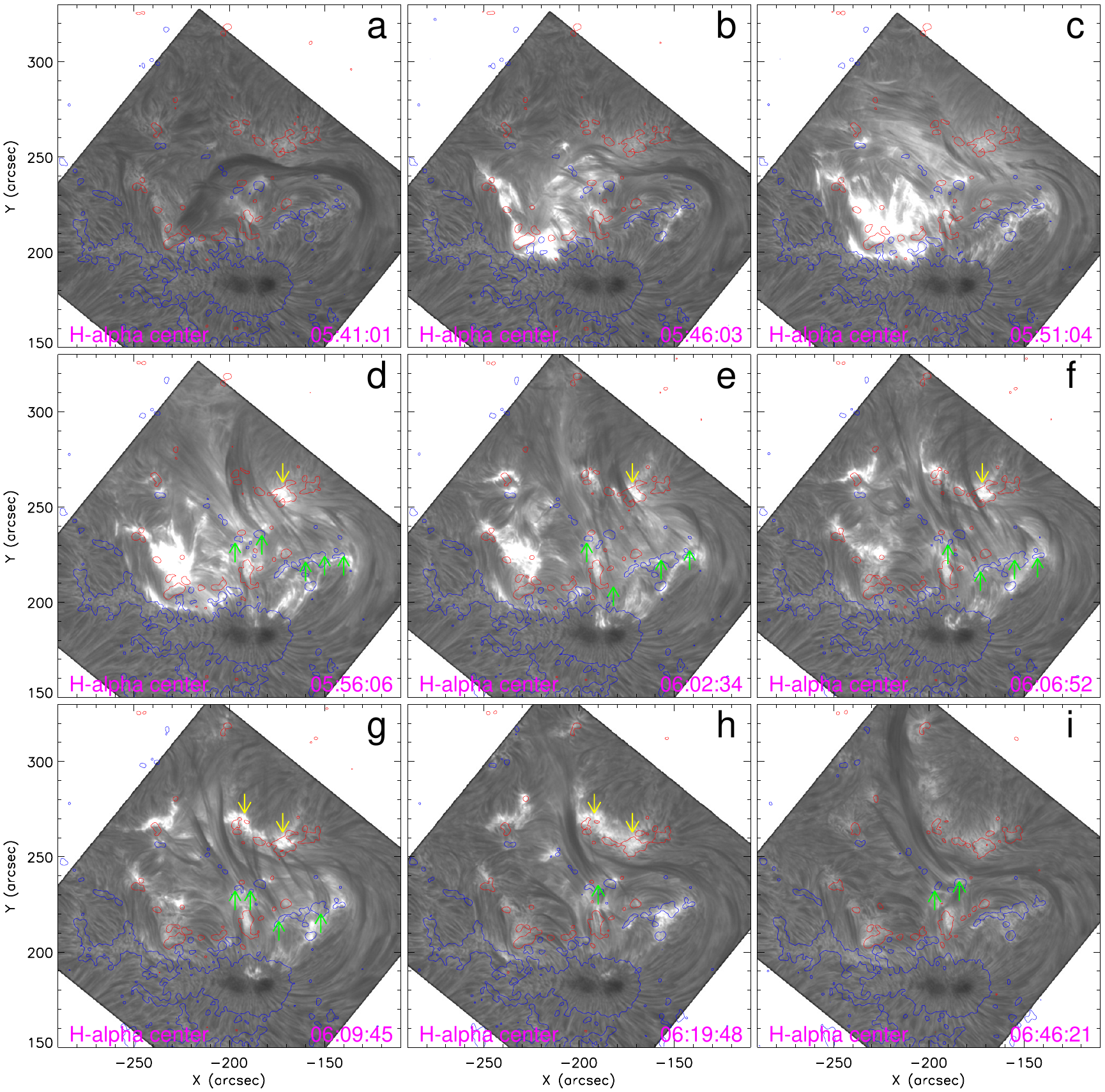}
\caption{Eruption process of the active-region filament in H$\alpha$ center images observed by the NVST with the contours of the line-of-sight magnetograms superimposed. The red and the blue contours represent the positive and the negative magnetic fields. The contour levels of magnetic fields are $\pm$ 100 G. The green arrows denote the foot-points of the filament and the yellow arrows denote the brightening sites at the north of the filament.  \textbf{An animation of the eruptive process is available in the online Journal. The animated H$\alpha$ center images include positive and negative magnetic field contours, and they proceed from 5:35 to 7:58 UT.} \label{fig3}}
\end{figure}

\begin{figure}
\plotone{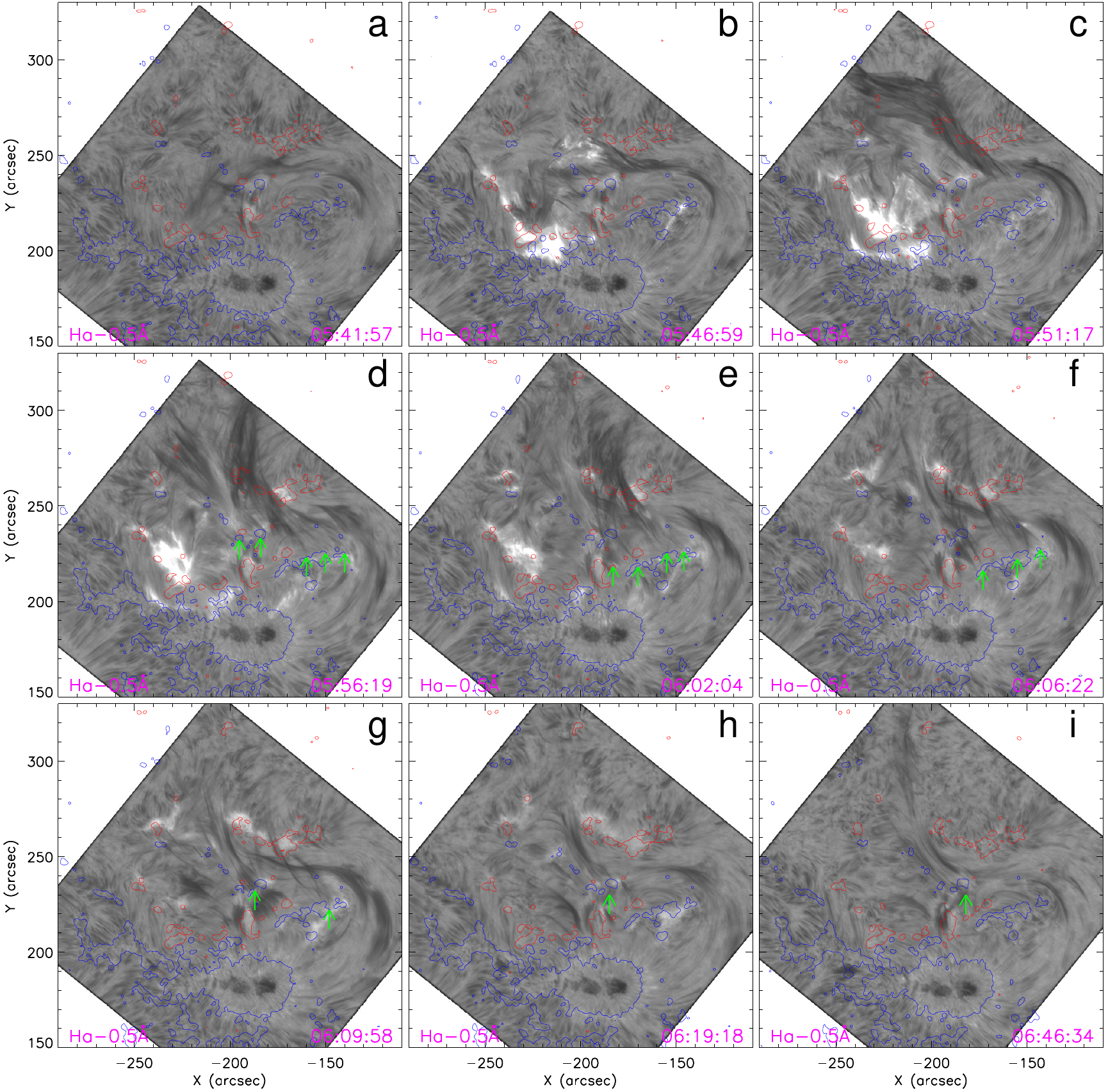}
\caption{Eruption process of the active-region filament in off-band H$\alpha$ center images (-0.5 \AA) observed by the NVST with the contours of the line-of-sight magnetograms superimposed. The red and the blue contours represent the positive and the negative magnetic fields. The contour levels of magnetic fields are $\pm$ 100 G. The green arrows denote the foot-points of the filament.  \label{fig4}}
\end{figure}

\begin{figure}
\plotone{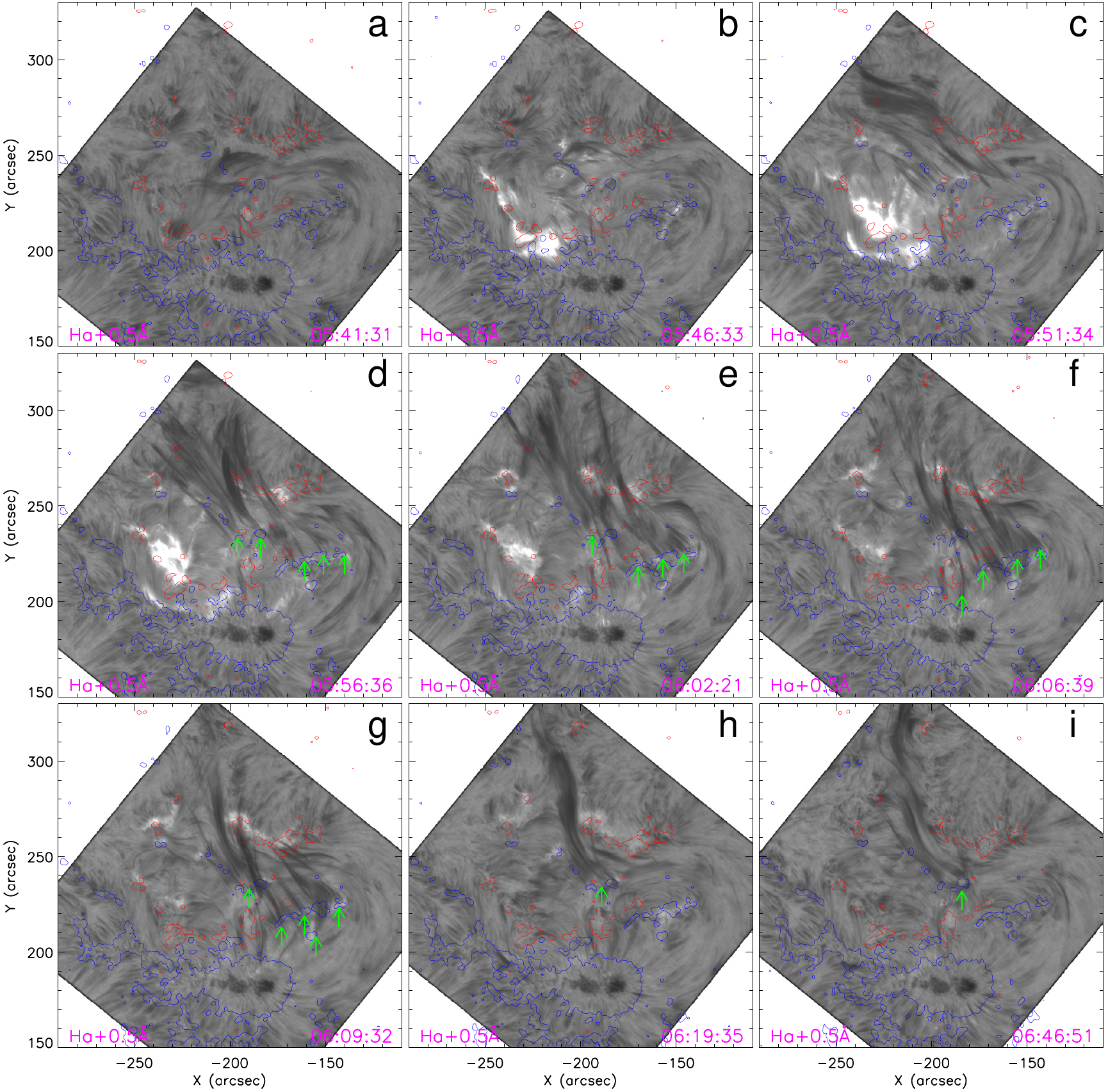}
\caption{Eruption process of the active-region filament in off-band H$\alpha$ center images (+0.5 \AA) observed by the NVST with the contours of the line-of-sight magnetograms superimposed. The red and the blue contours represent the positive and the negative magnetic fields. The contour levels of magnetic fields are $\pm$ 100 G. The green arrows denote the foot-points of the filament.  \label{fig5}}
\end{figure}

\begin{figure}
\plotone{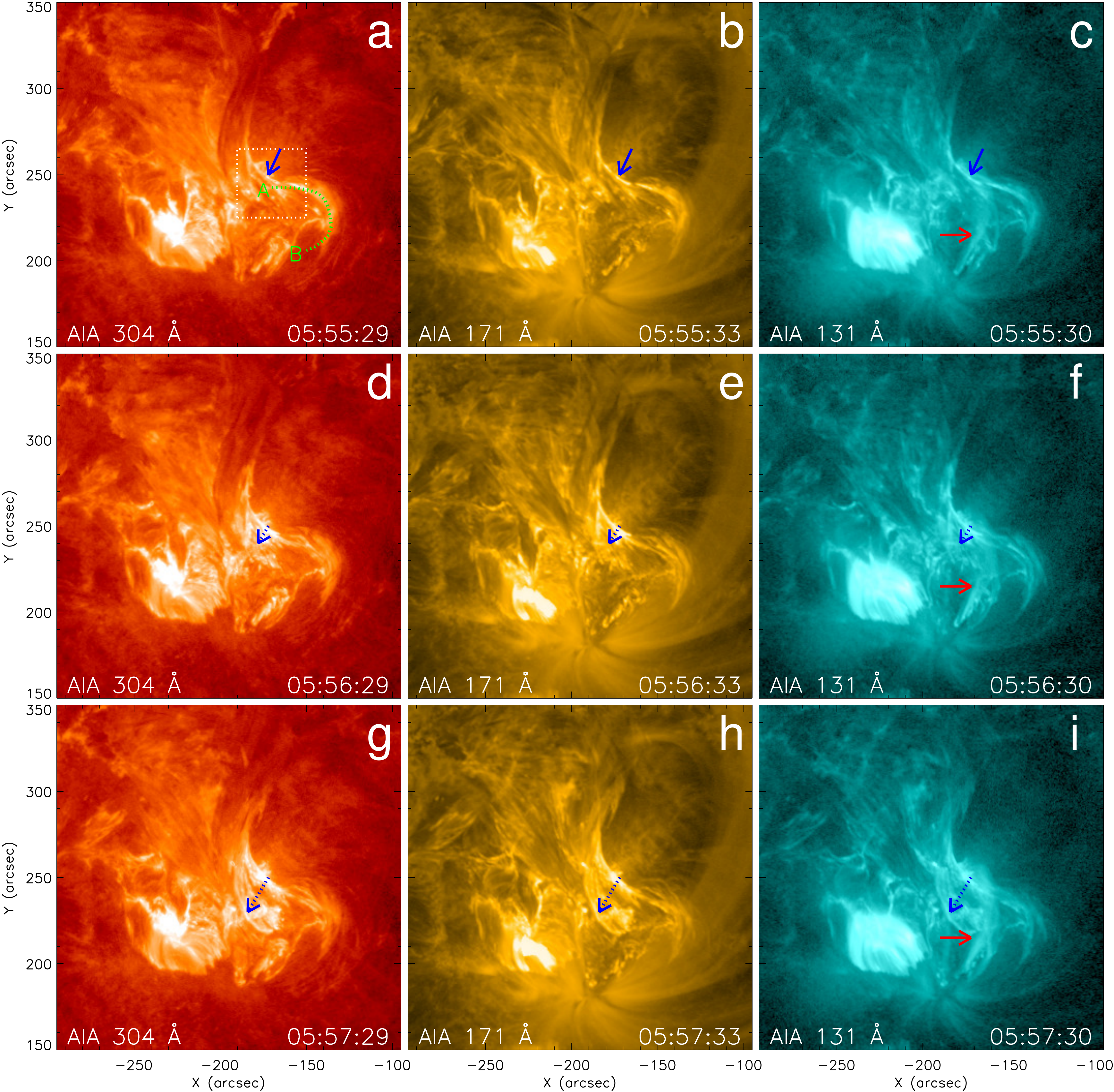}
\caption{Eruptive process of the filament shown in 304 \AA, 171 \AA, and 131 \AA\ images observed by the SDO/AIA. The blue arrows in the first row denote the original site of the brightening. The blue dotted arrows in Figs. 6d-6i denote the movement of the brightening from the right side to the left side of the filament. The red arrows denote the overlying magnetic loops seen in 131 \AA\ images. The box and the green dotted curve in Figure 6a show the region for calculating the change of intensity and the slit position of the time distance diagram of Figure 7. \textbf{An animation of this SDO/AIA images for the eruptive process of the filament is available in the online Journal. The animation covers a much longer duration than the static figure -- from 5:00 to 7:00 UT.} \label{fig6}}
\end{figure}

\begin{figure}
\plotone{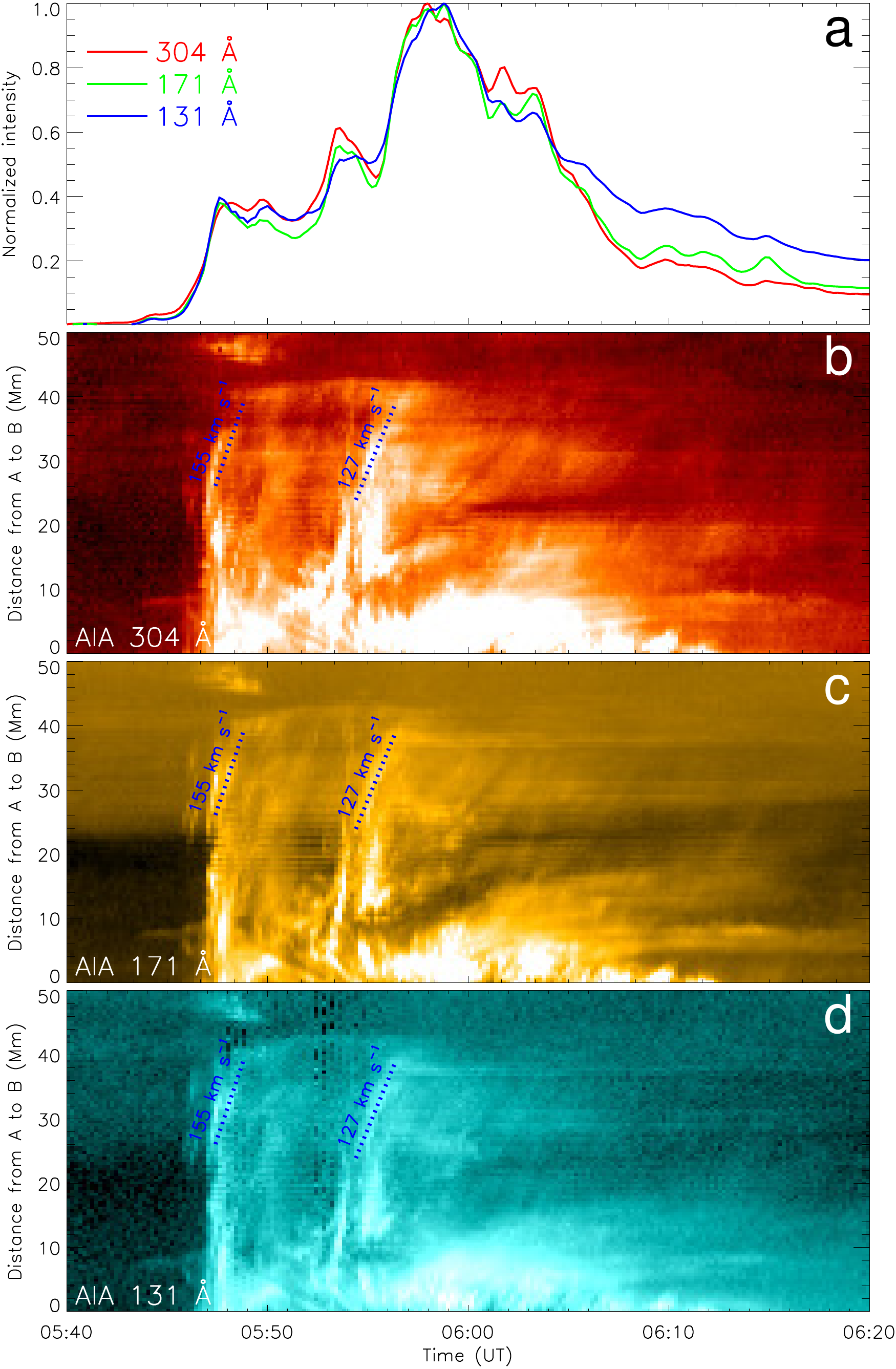}
\caption{(a): The profile of the normalized intensity of 304, 171, and 131 \AA\ images in the box in Figure 6a. (b-d): The time-distance diagrams along the green dotted line in Figure 6a by using 304, 171, and 131 \AA\ images. \label{fig7}}
\end{figure}

\begin{figure}
\plotone{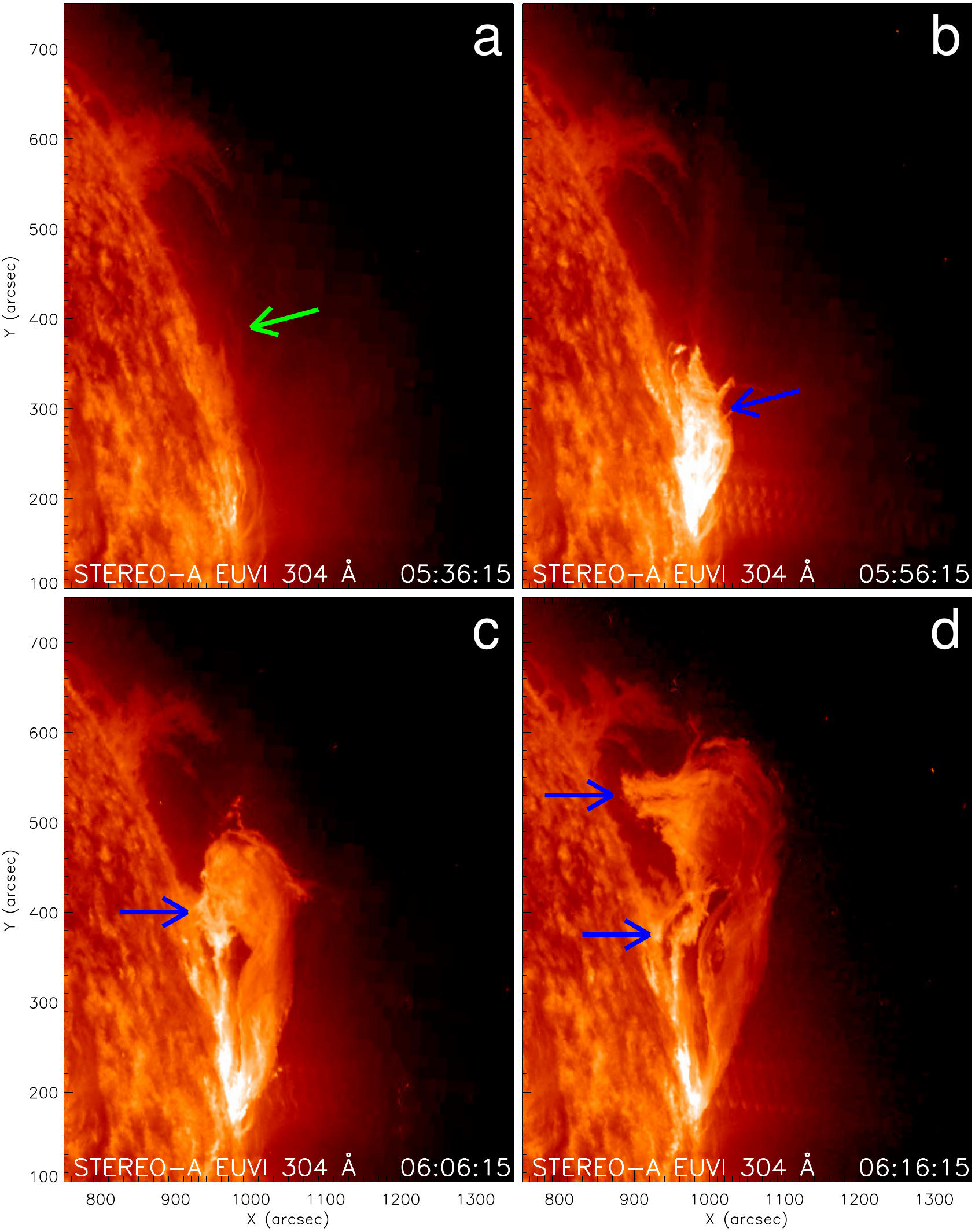}
\caption{Eruptive process of the filament shown in 304 \AA\ images observed by the STEREO A. The green arrow denotes the quiescent filament in the northwest of the filament. The blue arrow in Fig. 7b denotes the erupting filament. The blue arrows in Figs. 7c-7d denote the position where the filament materials fell down. \label{fig7}}
\end{figure}

\begin{figure}
\plotone{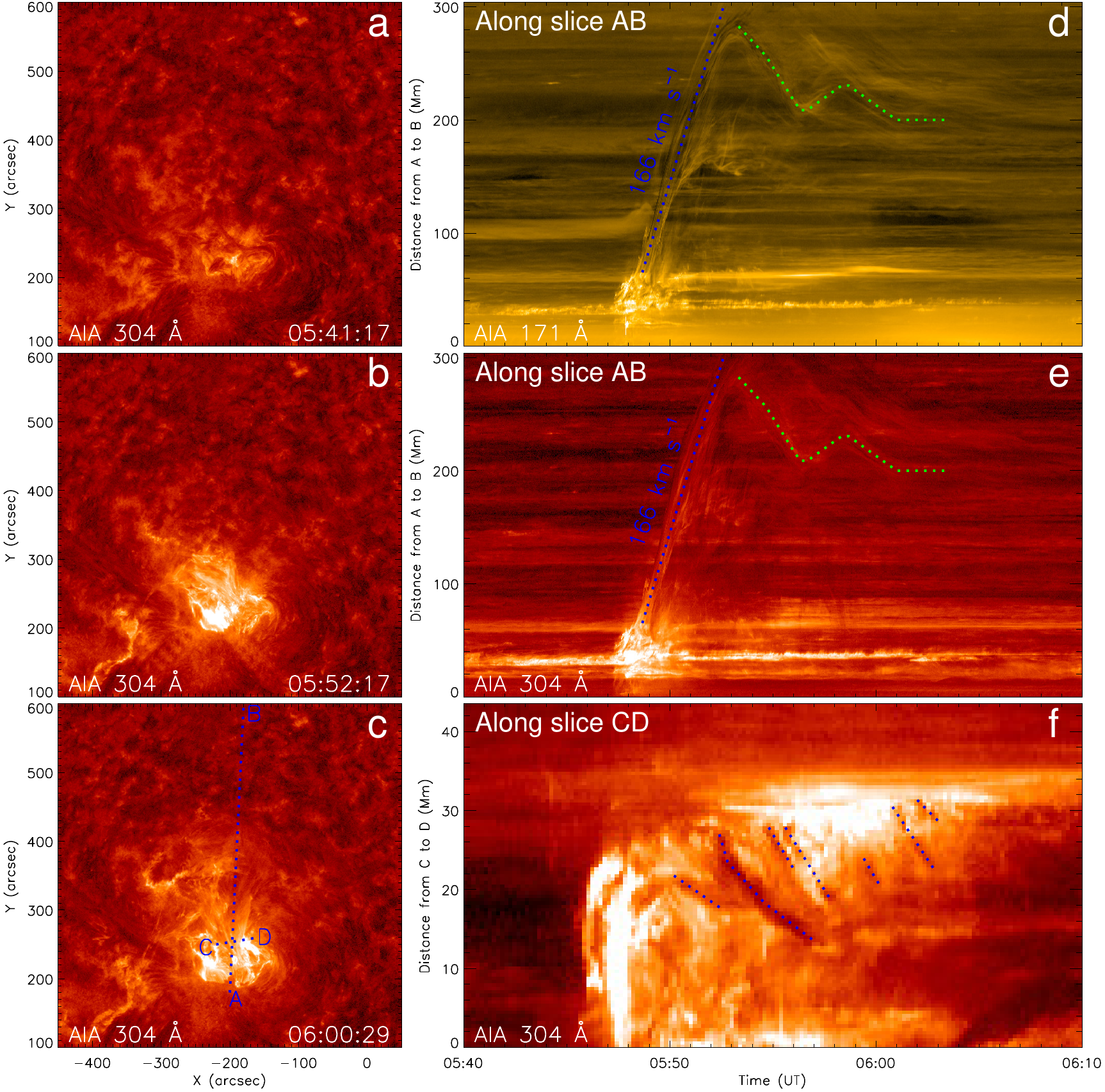}
\caption{Eruptive process of the filament associated with the C6.7 confined flare and time-distance diagrams along the slits AB and CD. (a)-(c) Evolution of the flare in 304 \AA\ images. The dashed blue lines AB and CD in Fig. 4c denote the positions of the time-distance diagrams of Figs. 4d, 4e and 4f. (d) Time-distance diagrams along the slit AB by using 171 \AA\ images. (e) Time-distance diagrams along the slit AB by using 304 \AA\ images. (f) Time-distance diagrams along the slit CD by using 304 \AA\ images. The dashed blue lines in Figs. 4a and 4b denote the eruption of the filament. The dashed green lines denote the oscillation of the filament structure during its eruption. The dashed line in Fig. 4c denote the untwisting motion of the filament threads during the filament eruption.  \label{fig6}}
\end{figure}

\begin{figure}
\plotone{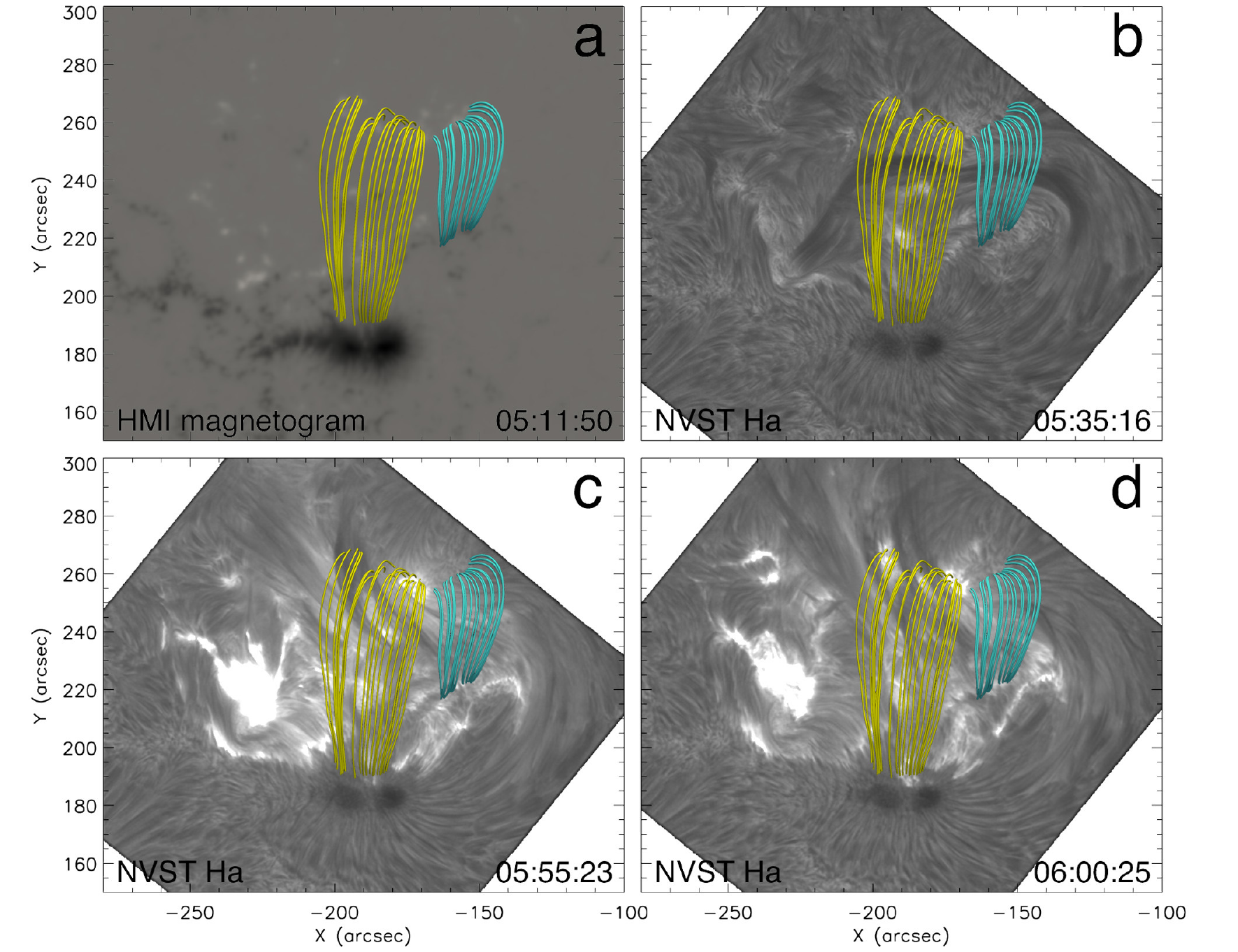}
\caption{The selected potential field lines in yellow and cyan superimposed on line-of-sight magnetogram at 05:11 UT  (a) and  H$\alpha$ images at 05:35 UT, 05:55 UT, and 06:00 UT (c-d). \label{fig7}}
\end{figure}

  \begin{figure}
  \centering
   \includegraphics[width=10cm]{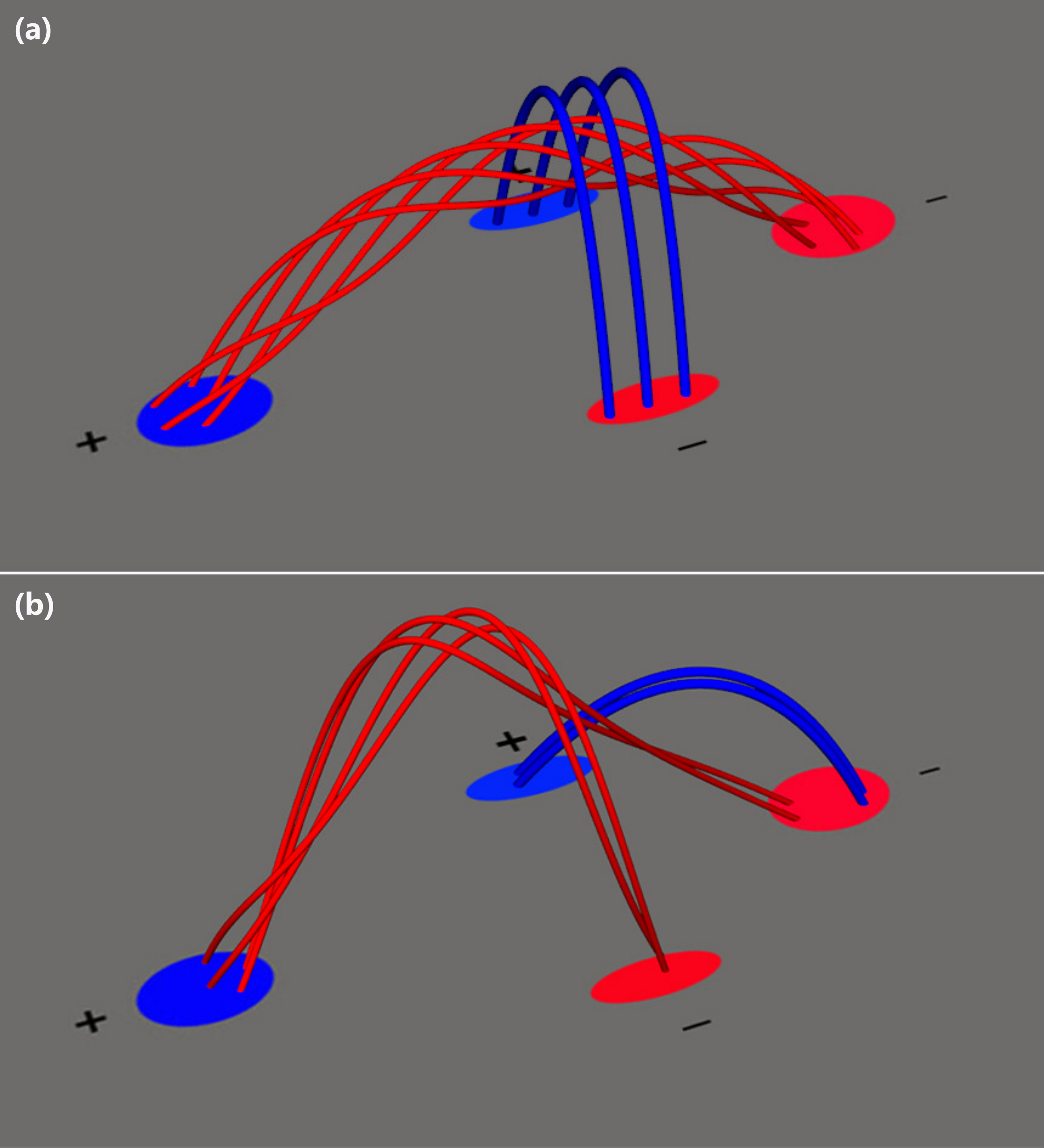}
\caption{Schematic sketch illustrating the change of the magnetic structures of the filament system before and after the filament eruption. The red and the blue patches indicate the negative and the positive polarities. The red curve lines indicate the magnetic structure of the filament. The blue closed loops indicate the overlying magnetic fields before the filament eruption in panel a and the newly formed magnetic loops.}
       \label{fig8}
   \end{figure}

\acknowledgments
The authors thank the referee for her/his constructive comments that helped to improve this paper and the NVST, SDO/ AIA, and SDO/ HMI teams for the high-cadence data support. This work is sponsored by the National Science Foundation of China( NSFC) under the grant numbers 11873087, 11603071,11503080, 11633008, by the Youth Innovation Promotion Association CAS (No.2011056) , by the Yunnan Science Foundation of China under number 2018FA001, by Project Supported by the Specialized Research Fund for State Key Laboratories, by Project Sup-ported by the Specialized Research Fund for State Key Laboratories and by the grant associated with project of the Group for Innovation of Yunnan Province. X.L.Y. thanks ISSI-BJ for supporting him to attend the team meeting led by J. C. Vial and P. F. Chen.

\end{document}